\documentclass[12pt]{article}
\usepackage{amssymb,amsmath}

\newcommand{\beq}[1]{\begin{equation}\label{#1}}
\newcommand{\nb}{{\not\mkern-3mu b}}
\newcommand{\np}{{\not\mkern-3mu\partial}}
\newcommand{\npp}{{\not\mkern-5mu p}}
\newcommand{\nk}{{\not\mkern-3mu k}}
\newcommand{\nA}{{\not\mkern-5mu A}}
\newcommand{\al}{{\alpha}}
\newcommand{\be}{{\beta}}
\newcommand{\de}{{\delta}}
\newcommand{\ga}{{\gamma}}
\newcommand{\mn}{{\mu\nu}}
\newcommand{\mna}{{\mu\nu\al}}
\newcommand{\ab}{{\al\be}}
\newcommand{\bm}{{b_\mu}}

\newcommand{\mnab}{{\mu\nu\al\be}}
\newcommand{\abmn}{{\al\be\mu\nu}}
\newcommand{\abgd}{{\al\be\ga\de}}
\newcommand{\mabg}{{\mu\al\be\ga}}
\newcommand{\cst}{Chern-Simons term}
\newcommand{\cs}{Chern-Simons}
\newcommand{\pv}{Pauli-Villars}
\newcommand{\dd}[2]{\mathop{\mathrm d\null}\nolimits^{#1}\!#2}
\DeclareMathOperator{\tr}{tr}
\newcommand{\fract}[2]{{\textstyle\frac#1#2}}
\let\eps\varepsilon
\let\bar\overline

\newcommand{\Section}[1]{\section{#1}\label{#1}\markright{\small                        
\thesection\enspace #1 }}
\begin{document}
\title{\bfseries When radiative corrections are finite\\
         but undetermined}
\author{R. Jackiw\thanks{MIT-CTP\#2835\quad
hep-th/9903044\qquad This work is partially supported by the US
Dept.~of Energy (DOE) under cooperative research agreement
DE-FC02-94ER40818.}\\
\small MIT Center for Theoretical Physics\\
\small Cambridge, Massachusetts}
\date{
\large\it Rajaramanfest, New Delhi, March 1999}
\maketitle
\begin{abstract}\noindent
In quantum field theory  radiative corrections can be finite but
undetermined. 
\end{abstract}
\thispagestyle{empty}

\setlength{\baselineskip}{1.1\baselineskip}

\Section{Introduction}

The following situation is familiar in quantum field
theory. One begins with a field-theoretic Lagrange
density that contains, in addition to its kinetic term, various
further contributions, which  at ``tree level''
describe masses of excitations and coupling strengths of
interactions. One might suppose that these take on definite
``classical'' values and  attempt to compute
corrections that arise when the fields are quantized. While the
hope is that such corrections are small -- $O(\hbar)$ -- mostly
one finds that they are infinite. It is then said that these
quantities -- masses and coupling strengths -- are not
calculable, and the classical plus (infinite) quantal
contributions are defined to take on definite, experimentally
determined values. In this way, the program of
renormalization, renormalization group, running coupling
constants, and so on, becomes an essential part of the theory. 

But this is not the only scenario. There are also favorable
situations where a tree-level value can be assigned; the
quantum correction is calculable and finite, giving a definite
(usually small) contribution. A physical example is the magnetic
moment of leptons, where the tree-level value of 2 for the
$g$-factor acquires precisely determined quantal corrections,
which these days are calculated and measured to six
significant figures.  A theoretical example is provided by the
Schwinger model -- two-dimensional QED with massless
fermions. The massless ``photon'' of the tree approximation
acquires at one loop (which is exact in this simple model) the
mass $e^2/\pi$, where $e$ is the coupling strength with
dimensionality of (mass)$^{1/2}$. [A similar effect is believed
to happen in three-dimensional
non-Abelian
 gauge theories, with radiatively induced mass $O(g)$, where
$g$ is the coupling strength with dimensionality of mass.]

In this essay, presented to Duggy Rajaraman on the occasion of
a significant birthday, I  call attention to a third
possibility. It can happen that radiative corrections are finite,
but not determined by the theory; so, just as for infinite
radiative corrections, their values are only fixed by
experiment. Rajaraman and I discussed an example of this 15
years ago~\cite{RJRR85}, and more recently another instance has been
encountered~\cite{RJVK99}.

\Section{Schwinger model}
Although the Schwinger model is simple, much studied, and
well understood, a close examination is useful for my further
discussion. The quantity of
interest is the effective action, radiatively induced by fermions:
\beq{eq:1}
\Gamma_{\mathrm S} (A) = -i \ln \det (i \np - e \nA)\ .
\end{equation}

Possible mass generation is seen in the $O(A^2)$ contribution,
which for this model determines~$\Gamma_{\mathrm S}
(A)$ completely.
To this end one computes the vacuum polarization, given in
momentum space by
\begin{subequations}\label{eq:2}
\beq{eq:2a}
\Pi^\mn_{\mathrm S} (p) = i \tr \int\frac{\dd2k}{(2\pi)^2}
\ga^\mu
\frac i\nk \ga^\nu \frac i{\nk+ \npp}\ .
\end{equation}
The integral is logarithmically divergent, hence shifting
integration variables does not alter its value. After the
divergent part is identified and separated, the convergent part
is evaluated, with the result
\beq{eq:2b}
\Pi_{\mathrm S}^\mn (p) = \tr \fract12\ga^\mu \ga_\alpha
\Pi^{\alpha\nu}_\infty + \fract1\pi \Bigl(\frac{g^\mn}2 -
\frac{p^\mu p^\nu}{p^2}
\Bigr)
\end{equation}
where
\beq{eq:2c}
\Pi_\infty^\mn \equiv 2i \int\frac{\dd2k}{(2\pi)^2}
\frac{(-k^2 g^\mn + 2k^\mu k^\nu)}{(k^2-\mu^2)^2} 
\end{equation}
\end{subequations}
and $\mu^2$ is an arbitrary infrared cutoff, whose value does not
affect $\Pi_\infty^\mn$.  Note that in general
$\Pi^\mn_{\mathrm S} (p)$ is not gauge invariant -- it is not
transverse to~$p_\mu$. In fact the tensor is traceless in
$(\mu,\nu)$, a feature already recognized in the integral
representation~\eqref{eq:2a} when it is remembered that in two
dimensions, $\ga^\mu \nk \ga_\mu = 0$. 

To make progress I must assign a value to $\Pi_\infty^\mn$.
But no unique value can be given, because the integral is
divergent, that is, undefined. By Lorentz invariance, 
$\Pi_\infty^\mn$ should be proportional to~$g^\mn$. In two
dimensions any Lorentz-invariant prescription for calculating
the integral will give a vanishing value, $\Pi_\infty^\mn=0$,
consistent with its being proportional to $g^\mn$ and traceless. 
Gauge invariance is regained by using, for instance,
Pauli-Villars regularization: the Pauli-Villars regulator fields
give an additional contribution
$
\Pi^\mn_{\mathrm{PV}} = \frac1{2\pi} g^\mn
$,
so that $\Pi^\mn_{\mathrm S} (p) +
\Pi^\mn_{\mathrm{PV}}$  is transverse. Alternatively,
an evaluation of
$\Pi^\mn_\infty$ by dimensional regularization gives
$\Pi^\mn_\infty =  \frac1{2\pi} g^\mn$, which also leads to a
gauge-invariant result for $ \Pi^\mn_{\mathrm S} (p)$. But I
shall not adopt these regularization procedures. The
viewpoint that I prefer will be used in the analysis of the models
considered below, for which the above regularization methods are
problematical.  

Since different evaluations of   $\Pi^\mn_\infty$ produce
different results, I propose that $\Pi^\mn_\infty = a g^\mn$,
where~$a$ is  dimensionless and as yet
\emph{undetermined}. This \emph{Ansatz} is consistent with
the fact that in $\Gamma_{\mathrm S}(A)$, terms cubic and higher
in~$A$ are well defined (actually they vanish), while the quadratic
term can have an undetermined, local contribution. 

Thus within my viewpoint, the Feynman graphs of the
Schwinger model need not be regulated, but they give a
vacuum polarization with an \emph{undetermined local part}:
\beq{eq:3}
\Pi_{\mathrm S}^\mn (p) = \fract1\pi \left(g^\mn
\Bigl(\frac{1+a}2\Bigr)
- \frac{p^\mu p^\nu}{p^2}
\right)\ .
\end{equation}

Now I make use of the formal gauge invariance of the
Schwinger model and enquire whether it is possible to fix the
ambiguity in~\eqref{eq:3} by insisting that this symmetry is
preserved: $\Pi^\mn_{\mathrm S}(p)$ should be transverse.
Indeed this is possible;  the choice $a=1$ yields the
conventional result for the vacuum polarization in this model
\beq{eq:4}
\Pi^\mn_{\mathrm S} (p) = \fract1\pi \Bigl( g^\mn -
\frac{p^\mu p^\nu}{p^2}\Bigr)
\end{equation}
and a photon mass
\beq{eq:5}
m^2 = \frac{e^2}\pi \ .
\end{equation}
Note that by adopting the transverse expression for the
vacuum polarization, in order to agree with the constraint of
gauge invariance, I have abandoned another formal feature:
tracelessness of~$\Pi^\mn_{\mathrm S} (p)$.  

I have passed slowly and laboriously over familiar ground so
that new territories can be explored quickly.  The key point is
that regularization has been avoided. While usual
regularization methods, for example, 
Pauli-Villars or dimensional, can be successfully used in the
Schwinger model, I shall now examine models for which these 
regularizations are unavailable or inappropriate. Nevertheless,
finite but undetermined radiative corrections can be calculated,  and
then further properties of the theory are brought to bear on the
question of whether or not the arbitrariness can be removed. 

\Section{First example: Chiral Schwinger model}

 My first example is the one that Rajaraman and I studied in
1985~\cite{RJRR85}: the chiral Schwinger model, where the vector
interaction of the Schwinger model is replaced by a chiral interaction
$e(1+\ga_5)\nA$, $\ga_5=\ga^0 \ga^1$.
The relevant induced action now reads
\beq{eq:6}
\Gamma_{\mathrm{CS}} (A) = -i \ln \det \bigl(i \np -
e(1+\ga_5) \nA\bigr)\ .
\end{equation}
Evaluation of the vacuum polarization proceeds as above in the
Schwinger model to the end
\beq{eq:7}
\Pi_{\mathrm{CS}}^\mn (p) = \fract1\pi 
\Bigl(
g^\mn a - (g^{\mu\alpha}+\eps^{\mu\alpha}) \frac{p_\alpha
p_\beta}{p^2} (g^{\beta\nu}-\eps^{\beta\nu})
\Bigr)\ .
\end{equation}
Here $a$ is once again a dimensionless parameter, not
determined uniquely by the different procedures for 
calculating the fermion determinant; it gives a local,
$O(A^2)$, contribution to $\Gamma_{\mathrm{CS}} (A)$. Note
further that since the usual Schwinger model is the
sum of two chiral models with opposite chirality,
combining~\eqref{eq:7} with its chiral partner ($\eps^\mn \to
-\eps^\mn$) reproduces formula~\eqref{eq:3}
for~$\Pi_{\mathrm S}^\mn(p)$. 

Can a regulator method be implemented here, as it can be for
the Schwinger model, to remove the ambiguity? Pauli-Villars
regulators with massive fermions are obviously inappropriate
because the chiral interaction requires massless fer\-mi\-ons.
Dimensional regularization is problematic with a $\ga_5$
matrix, which is dimension specific. So there only remains the
possibility of enforcing gauge invariance -- a formal property
of the theory. 

However, unlike for the Schwinger model, imposing
transversality on $\Pi_{\mathrm{CS}}^\mn (p)$ does not
determine~$a$, because the longitudinal part does not vanish
for any value of~$a$:
\beq{eq:8}
p_\mu \Pi_{\mathrm{CS}}^\mn (p) =
\fract1\pi \bigl(p^\nu (a-1) + p_\mu \eps^\mn\bigr)\ .
\end{equation} 
This of course is another face of the two-dimensional chiral
anomaly~\cite{RJ85} -- owing to the anomalous nonconservation of
the chiral current, the quantized chiral Schwinger model is not gauge
invariant. Nevertheless, it possesses a physical spectrum for 
$a>1$, with radiatively induced photon mass~$m$:
\beq{eq:9}
m^2 = \frac{e^2}\pi \frac{a^2}{a-1}\ . 
\end{equation}
Thus here radiative corrections are finite but undetermined, so
that if a physical setting for this model can be found (perhaps
in a description of edge states in the quantum Hall effect), the
value of~$a$ and~$m$ is fixed only by experiment. 

Finally we note that for $a\neq0$, a formal property of the
chiral Schwinger model is abandoned: owing to the
two-dimensional identity $\ga^5\ga^\mu = \eps^\mn
\ga_\nu$, the gauge field $A_\mu$ enters~\eqref{eq:6}
only on the combination $(g^\mn + \eps^\mn)A_\nu$, which is
consistent with the unique, absorptive part of~\eqref{eq:7},
but not with the real part. [In the present context, this
corresponds to abandoning the tracelessness
of~$\Pi_{\mathrm S}^\mn(p)$ for the vector Schwinger model.]

\Section{Second example: Triangle graphs}

My second example of finite  but undetermined  radiative
corrections is even older -- I recall the massless, fermionic
triangle-loop graphs in four dimensions with vector, vector, and axial
vector vertices: $\Gamma^{\mna} (p,q)$. (The incoming vector
momenta are $p^\mu$ and $q^\nu$, while the outgoing axial vector
momentum is $p^\alpha + q^\alpha$~\cite{RJ85,SA70}.)  Because
three fermion propagators determine the triangle, the Feynman
graphs are (superficially)  linearly divergent (even though an
eventual evaluation,  relying on a Lorentz-invariant calculation,
yields a finite answer). However, owing to the linear divergence,
shifting the integration momentum in the closed loop changes the
value of the integral, so that there is an essential ambiguity
in $\Gamma^{\mna} (p,q)$: an evaluation of the integral
produces some preferred form, plus an undetermined
contribution proportional to $\eps^{\abmn}
(p-q)_\nu$:
\beq{eq:10}
\Gamma^\mna (p,q) \sim \Gamma^\mna (p,q) +
ia\eps^\mnab(p-q)_\be\ .
\end{equation}
Here $a$ is  a dimensionless constant, controlling the
magnitude of an arbitrary local part. Turning to  
symmetries/formal properties to fix~$a$, I try to make use of
the conservation of the vector current and (since fermions are
massless) of the axial vector current, thereby requiring
transversality of $\Gamma^\mna(p,q)$ in each index. But as
is well known, for no value of~$a$ can this condition be
satisfied, and this is another face of the four-dimensional
chiral anomaly. The situation is completely analogous to the
chiral Schwinger model. So we must abandon some of the
formal properties, and settle for transversality in the vector
indices or in the axial vector, but not in all three. Moreover, the
calculation of the radiative correction cannot decide which
option to choose -- this must come from elsewhere in the
theory. In other words, the ``correct'' answer for the triangle
graph is not intrinsic to it, but depends on the context in
which it arises. Thus, for example, when the vector indices
couple to photons and refer to gauge currents, while the axial
vector refers to a global chiral symmetry, the choice is made
to preserve transversality of the vector indices and to
abandon axial-vector transversality. This is the situation for
$\pi^0\to 2\ga$ decay~\cite{JBRJ69}. On the other hand, in the
standard model of particle physics, when axial vertices are part
of the chiral coupling to gauge fields and a vector index refers to
a global fermion-number current, transversality of the former
rather than the latter is enforced. This is the situation with
't~Hooft's celebrated calculation of proton decay in the standard
model~\cite{GtH76}.

I must emphasize that both
Pauli-Villars and a specific dimensional regularization~\cite{MPDS95}
preserve vector gauge invariance. Nevertheless, as explained above,
this need not be the correct choice if chiral invariance should be
enforced. 

Note that once a decision is made about which symmetry
(transversality) should be preserved, a unique value for
$\Gamma^\mna (p,q)$ is established. An arbitrary value persists only
when no symmetry is enforced.

\section{Third example: Induced Lorentz-PTC\protect\newline symmetry
breaking}\label{sec:5} 
\markright{\small \thesection\enspace Third example: Induced
Lorentz-PTC symmetry breaking}

It is known that if one adds to conventional four-dimensional
Maxwell electrodynamics the Lorentz- and PTC-violating
\cst\ 
\beq{eq:11}
\Delta{\cal L} = \fract12 c_\mu \eps^\mabg F_\ab A_\ga
\end{equation}
where $c_\mu$ selects a fixed direction in spacetime, light
from distant galaxies undergoes a Faraday-like
rotation~\cite{SCGFRJ90}. Observation of distant galaxies puts a
stringent bound on this ``vacuum birefringence'': 
$c_\mu$ should effectively vanish~\cite{SCGFRJ90,MGVT96}. An
important feature of the
\cst\  is that its Lagrange density is not gauge invariant:
$\Delta\cal L$ depends on~$A_\mu$. However, the action, the
spacetime integral of the density, is gauge invariant because
under a gauge transformation $\Delta\cal L$ changes by a
total derivative. Correspondingly, the Euler-Lagrange equations
remain gauge invariant, even in the presence of the 
gauge-noninvariant \cs\ density~\cite{SDRJST82}. 

A natural question is whether such a term could be induced through
radiative corrections when Lorentz and PTC symmetries are violated in
other parts of a larger theory. 

For definiteness, consider the fermionic Lagrange density of
conventional QED, extended to include a Lorentz- and
PTC-violating axial vector term~\cite{DCVK98,SCSG98,JMCPO98}
\beq{eq:12}
{\cal L}_{\mathrm{extended}} = \bar\psi \bigl(i\np - e\nA - m - \nb
\ga_5\bigr) \psi
\end{equation}
where  $\ga_5$ is Hermitian with $\tr \ga_5
\ga^\al
\ga^\be \ga^\ga
\ga^\delta = 4 i\eps^\abgd$
 and $\bm$ is a constant, prescribed 4-vector.  One is then led to
enquire whether the effective action
\beq{eq:13}
\Gamma (A) = -i\ln\det \bigl(i\np - e\nA - m - \nb
\ga_5\bigr) 
\end{equation}
contains the \cst\  \eqref{eq:11} with $c_\mu$ determined by
$\bm$. Since \eqref{eq:11}  is bilinear in $A_\mu$, I examine only
the
$O(A^2)$ portion of~$\Gamma(A)$. 

A plausible approach is to calculate the lowest order in~$\bm$.
But then it is clear that one again encounters triangle graphs,
with two vector vertices coupling to the two $A_\mu$'s and the
axial vertex contracted with~$\bm$. Moreover, the axial vector
carries zero momentum: in the notation of the previous
subjection $p=-q$, and the relevant amplitude is
$b_\al \Gamma^\mna (p,-p)$ where now $\Gamma^\mna (p,q)$
is calculated with massive fermions. The coefficient of the
induced {\cst}s is determined by $b_\al \Gamma^\mna (p,-p)
\bigr|_{p^2=0}$.

It follows from the property of the triangle graphs, which I reviewed
previously, that $\Gamma^\mna (p,-p)$ is undetermined as
in~\eqref{eq:10}:
\beq{eq:14}
\Gamma^\mna (p,-p) \sim \Gamma^\mna (p,-p) +
2ia\eps^\mnab p_\be\ .
\end{equation}
Of course   gauge invariance, that is, transversality of
$\Gamma^\mna (p,-p)$ to $p_\mu$ and $p_\nu$, must hold.  But
unlike the case when the axial vector carries nonvanishing
momentum, 
 $\Gamma^\mna (p,-p)$ is
transverse in its photon indices for all routings of the integration
momentum, as is also seen from the form of the ambiguity:
$\eps^\mnab p_\be$ is transverse to $p_\mu$ and~$p_\nu$.
Therefore, the requirement of gauge invariance does not fix a
value for the graphs in the present context. 

The conclusion is that the radiatively induced \cst, to lowest
order in $\bm$, is as in \eqref{eq:11}, with $c_\mu$
proportional to
$\bm$, but the numerical proportionality constant is
undetermined when the theory is considered perturbatively
in~$\bm$~\cite{RJVK99,DCVK98}. 

What is the situation with regulators? \pv\ regularization removes 
the induced \cst. This is true because the induced undetermined
coefficient is a mass-independent pure number, and an equal
amount is subtracted by
\pv\ regulators. Dimensional regularization is problematic in the
presence of a $\ga_5$ matrix, and a variety of results can be obtained
when a variety of prescriptions is made for the dimensional
generalization of~$\ga_5$. One can arrange a dimensional
extension~\cite{MPDS95} so that the induced \cs\ coefficient
vanishes. 

However, these regulators are inappropriate for the following subtle
reason~\cite{RJVK99}. We are seeking an induced density,
\eqref{eq:11}, which is not gauge invariant, while its spacetime 
integral is gauge invariant. An alternative, equivalent
momentum-space statement is that we seek a term that is gauge
invariant at zero momentum, but not at arbitrary momenta.
\pv\ regularization and gauge-invariant dimensional
regularization enforce gauge invariance at all momenta, and
therefore exclude
\emph{a~priori} the generation of a \cst, which does not have
this property.  These regulators are not 
sufficiently fine to enforce gauge invariance
at
 zero momentum only. Thus an  undetermined but finite answer
remains.

Is there any other criterion that can be brought to bear on the
problem? Coleman and Glashow~\cite{SCSG98} have suggested that
the axial vector current be regulated so  it is gauge invariant at
all momenta. This gives a unique result for the induced \cst\ -- it
vanishes. However, as stated above, such a regularization
requirement (which can be implemented by  \pv\ or
dimensional procedures~\cite{MPDS95}) is
\emph{ad~hoc} and unjustified since the axial vector enters into the
theory only at zero momentum (it couples to an external constant,
$\bm$).  Demanding gauge invariance for all momenta is equivalent
to demanding gauge invariance for the unintegrated densities, and
this would exclude \emph{a~priori} the \cst, which is not a
gauge-invariant density. 

Curiously,  a unique answer does emerge when the interaction
$\bar\psi\nb\ga_5 \psi$ is treated nonperturbatively. Note that the
$\bm$-exact,
$O(A^2)$ contribution to the effective action is determined by the
vacuum polarization tensor constructed with $\bm$-exact fermion
propagators
\beq{eq:15}
\Pi^\mn (p) = i\tr \int \frac{\dd4k}{(2\pi)^4} \ga^\mu \frac i{\nk -m
-\nb \ga_5} \ga^\nu \frac i{\nk + \npp  -m -\nb \ga_5} \ .
\end{equation} The  $\bm$-exact propagator 
\beq{eq:16} G(p) =  \frac i{\nk -m -\nb \ga_5} 
\end{equation} may be decomposed as
\beq{eq:17} G(p) =   S(p) - i G(p) \nb \ga_5  S(p)\equiv S(p) + G_b
(p)
\end{equation} where $S(p)$ is the free propagator
\beq{eq:18} S(p) =  \frac i{\npp - m} \ .
\end{equation} With decomposition $\Pi^\mn (p)$ splits into three
terms
 \beq{eq:19}
 \Pi^\mn = \Pi^\mn_0 + \Pi^\mn_b  + \Pi^\mn_{bb}\ .
 \end{equation}
$\Pi^\mn_0$ is the usual, lowest-order vacuum polarization tensor of
QED, which I shall not discuss further. $\Pi^\mn_{bb}$, containing
$G_b$ twice, is at least quadratic in $b$; it is at most logarithmically
divergent and suffers no ambiguities in routing the internal
momenta. The $\bm$-linear contribution to the induced \cst\ arises
from $\Pi^\mn_b$, which is explicitly given~by
 \beq{eq:20}
 \Pi^\mn_b  (p)  = i\tr \int \!\frac{\dd4k}{(2\pi)^4} \Bigl\{
\ga^\mu S(k) \ga^\nu G_b(k+p) + \ga^\mu G_b(k) \ga^\nu S(k+p) 
\Bigr\}\ .
 \end{equation} 
There are several important features of this
expression. Each of the two integrals is (superficially) linearly
divergent. However, the divergences cancel when the two terms are
taken together and the traces are evaluated. As a consequence, there
is no momentum-routing ambiguity in the summed integrand. When
the integration momentum is shifted by the same amount on
\emph{both} integrands, the value of the integral does not change,
even though shifting separately by different amounts in each of the
two integrands changes the value of the integral by a surface term. It
follows that different momentum routings in the entire
integrand~\eqref{eq:20} leave the value of $\Pi^\mn_b$ unchanged,
because they produce a simultaneous shift of integration variable by
the same amount in each of the two integrands. Therefore, a unique
value can be attached to~$\Pi^\mn_b(p)$.

Next we evaluate $\Pi^\mn_b(p)$ to lowest order in $\bm$, 
by replacing $G_b(k)$ with\linebreak 
 $-i S(k) \nb \ga_5 S(k)$. This gives
$\Pi^\mn_b(p)
\simeq
b_\al\Pi^\mna (p)$, where
 \beq{eq:21}\begin{split}
 \Pi^\mna  (p)  =  \tr \int\!\! \frac{\dd4k}{(2\pi)^4} &\Bigl\{
\ga^\mu S(k) \ga^\nu S(k+p) \ga^\al\ga_5 S(k+p)\\ &\quad{} +
\ga^\mu S(k) \ga^\al \ga^5 S(k) \ga^\nu S(k+p)
\Bigr\}\ . \end{split}
 \end{equation} 
When Lorentz invariance is enforced, this integral may be evaluated
unambiguously. It leads to the finite result
\beq{eq:22}\begin{split}
 \Pi^\mna  (p)  &= -i \eps^\mnab p_\be \Bigl(
\frac1{8\pi^2} + \frac2{\pi^2} \int_{2m}^\infty\!\! \dd{}\sigma
\frac{m^2}{\sqrt{\sigma^2-4m^2}} \frac1{p^2-\sigma^2-i\eps}
\Bigr)\\ &= i\eps^\mnab \frac{p_\be}{2\pi^2}
\Bigl(\frac\theta{\sin\theta} -
\frac14\Bigr)
\end{split}
 \end{equation} where $\theta=2\sin^{-1}\sqrt{p^2}/2m$, so that
\begin{subequations}\label{eq:23}
\beq{eq:23a}
 \Pi^\mna  (p) \Bigr|_{p^2=0} = \frac{3i}{8\pi^2} \eps^\mnab p_\be
\end{equation}  and the \cs\ coefficient is
unambiguously~\cite{RJVK99,JMCPO98}
\beq{eq:23b} c^\mu = \frac3{16\pi^2} b^\mu\ . 
\end{equation}
\end{subequations}

What is the difference between the present calculation and the
perturbative ones that leave $c^\mu$ undetermined?
Eq.~\eqref{eq:21} does indeed exhibit the vector, vector, axial vector
triangle graphs, with zero 4-momentum in the axial vertex, as in the
$\bm$-perturbative calculation. However, here the two triangle
graphs have their integration momenta routings correlated, since they
descend from the single,
$\bm$-exact formula~\eqref{eq:15}.  In $\bm$-perturbation theory,
no correlation is determined \emph{a~priori} between the momentum
routings of the two graphs. If in the perturbative calculation the
relative routings are as in~\eqref{eq:21}, the resulting expression
coincides with~\eqref{eq:22}. Otherwise a shift of integration
variables in one integrand relative to the other produces the
configuration~\eqref{eq:21}, but generates a surface contribution.
Therefore, a $\bm$-perturbative calculation gives 
\eqref{eq:22} with an undetermined polynomial contribution
\beq{eq:24}\begin{split}
 \Pi^\mna  (p)  &= -i \eps^\mnab p_\be \Bigl(
\frac{a}{2\pi^2} + \frac2{\pi^2} \int_{2m}^\infty\!\! \dd{}\sigma
\frac{m^2}{\sqrt{\sigma^2-4m^2}} \frac1{p^2-\sigma^2+i\eps}
\Bigr)\\ &= i\eps^\mnab \frac{p_\be}{2\pi^2}
\Bigl(\frac\theta{\sin\theta} - a\Bigr)
\end{split}
 \end{equation} leading to a \cs\ density with strength
\beq{eq:25} c^\mu = \frac1{4\pi^2} (1-a) b^\mu \ .
 \end{equation} Note that gauge invariance is preserved for all~$a$,
in the sense that
$p_\mu \Pi^\mna (p)=0$. (In momentum space one does  not see the
gauge noninvariance of the position-space \cs\ density.)

As in our other examples, the arbitrary term is a local contribution to
the effective action \eqref{eq:13} and in a dispersive representation,
as in~\eqref{eq:24}, it contributes a real subtraction, even though the
unique absorptive part permits presenting an unsubtracted integral 

An open question remains whether the unique result obtained by
the nonperturbative evaluation has any deeper significance. 

[The demand that triangle graphs be evaluated so that they are
gauge invariant with arbitrary momentum in the  axial vector
vertex~\cite{SCSG98},  can be met by a particular routing of
the integration momentum~\cite{RJVK99,RJ85} or by Pauli-Villars
regularization, or by a particular dimensional
regularization~\cite{MPDS95}; one finds $a=1$ and $c^\mu=0$.
But there is no \emph{a~priori} reason for placing this
requirement on the theory, and one may view its consequence
($c^\mu=0$) as  tautological, since a \cst\ does not enjoy such a
strong form of gauge invariance.]

Experiment effectively prohibits a \cst\ in QED, and I 
conclude that the conjectural noninvariant addition to fermion
dynamics,
$\bar\psi\nb\ga_5\psi$, is absent altogether (if the nonperturbative
approach is taken) or that the ambiguity is fixed so that $c_\mu=0$
(if the undetermined perturbative calculation is taken).  When several
different fermion species participate in the Lorentz- and
PTC-violating interaction, another possibility emerges: contributions
from different fermions may sum to zero~\cite{DCVK98}.

\Section{Conclusion}

The various radiative corrections that I surveyed ($g-2$, induced
mass in the Schwinger and chiral Schwinger models, triangle graphs
of the chiral anomaly, induced Lorentz and PTC-violating \cst) 
behave variously in their calculability. Can one formulate a criterion
that will settle \emph{a~priori} whether the radiative correction
produces a definite or indefinite result? The above examples suggest
the following rule: If the form of the radiative correction is such that
inserting it into the bare Lagrangian would interfere with symmetries
of the model or  would spoil renormalizability, then the radiative
result will be finite and uniquely fixed. Alternatively, if modifying the
bare Lagrangian by the radiative correction preserves
renormalizability and retains  the symmetries of the theory, then the
radiative calculation will not produce a definite result -- it is as if the
term in question is already present in the bare Lagrangian with an
undetermined strength, and the radiative correction adds a further
undetermined contribution. With this criterion, the radiatively
induced  $g-2$ Pauli term in QED and the photon mass in the
Schwinger model are unique, because in the former case a Pauli term
in the bare Lagrangian would spoil renormalizability, and in the
latter, a photon mass term would spoil gauge invariance.
Correspondingly, in the quantified chiral Schwinger model, the chiral
anomaly spoils gauge invariance, so there is no symmetry prohibiting
inserting a photon mass in the bare Lagrangian, and indeed the
radiatively induced mass is undetermined. Similarly with the
symmetry-violating \cst: ignoring experimental constraints and
allowing symmetry violation in the fermion sector permits insertion
of a \cst\ in the Lagrange density, so the radiatively induced \cst\
can have an arbitrary coefficient. (The subtlety here is that one must
allow for calculational schemes that permit  gauge noninvariance on
the level of a position-space density.) Finally, the triangle graphs of
the axial vector anomaly take on a unique value once it is decided
which symmetries are preserved, vector or chiral. In either case they
induce a process that, if inserted as a vertex in the bare Lagrangian,
would destroy renormalizability: the induced $\pi^0$ decay vertex is
$\pi^0 \eps^\mnab F_\mn F_\ab$ and the induced baryon decay
amplitude involves high powers of Fermi fields. Thus the radiatively
induced values are unique. 

When a radiative correction is infinite, there must be a term in the
bare Lagrangian of similar form to absorb the infinity through
renormalization. The finite but undetermined radiative corrections,
which I have discussed, are seen to be similar to the infinite ones, and
their finite value does not enhance the predictability of the theory.

\end{document}